\documentclass[10pt, twocolumn, twoside]{article}

\usepackage[utf8]{inputenc}
\usepackage[T1]{fontenc}
\usepackage{lmodern}
\usepackage{microtype}          

\usepackage[
  top=1in, bottom=1in,
  left=0.75in, right=0.75in,
  columnsep=0.25in
]{geometry}
\usepackage{multirow}

\usepackage{amsmath, amssymb, amsthm}
\usepackage{mathtools}
\usepackage{gensymb}
\usepackage{bbm}

\usepackage{graphicx}
\usepackage{float}
\usepackage[font=small, labelfont=bf]{caption}
\usepackage{subcaption}

\usepackage{booktabs}
\usepackage{array}

\usepackage{algorithm}
\usepackage{algpseudocode}

\usepackage[authoryear, round]{natbib}

\usepackage[
  colorlinks=true,
  citecolor=blue,
  linkcolor=blue,
  urlcolor=blue
]{hyperref}
\usepackage{orcidlink}
\usepackage{hyperref} 
\usepackage{cleveref}

\usepackage{xcolor}
\usepackage{lipsum}               
\usepackage{balance}              

\usepackage{fancyhdr}
\fancypagestyle{plain}{%
  \fancyhf{}%
}

\title{%
  \textbf{Capturing Aleatoric Uncertainty in Climate Models}
}

\author{%
  Cornelia Gruber \orcidlink{0009-0002-0657-3558} \textsuperscript{*1} \quad
  Henri Funk \orcidlink{0009-0007-0949-8385} \textsuperscript{*1,2,3} \\
  Magdalena Mittermeier \orcidlink{0000-0002-8668-281X} \textsuperscript{2}\quad
  Helmut Küchenhoff \orcidlink{0000-0002-6372-2487} \textsuperscript{1}\quad
  Göran Kauermann \orcidlink{0000-0003-0742-7835} \textsuperscript{1,3} \\[4pt]
  \textsuperscript{1}Department of Statistics, LMU Munich, Munich, Germany\\
  \textsuperscript{2}Department of Geography, LMU Munich, Munich, Germany\\
  \textsuperscript{3}Munich Center for Machine Learning, MCML, Munich, Germany\\[4pt]
  \textsuperscript{*} contributed equally, \quad Correspondence: \texttt{cornelia.gruber@lmu.de}
}

\date{}   

\begin{document}

\twocolumn[{%
  \maketitle
  \thispagestyle{plain}

  \begin{center}
    \begin{minipage}{0.92\linewidth}   
      \begin{abstract}
        \noindent
        Internal climate variability arises from the climate system’s inherently chaotic dynamics.
Quantifying it is essential for climate science, as it enables risk-based decision-making and differentiates between externally forced change and internal fluctuations.
In statistical terms, natural variability corresponds to aleatoric uncertainty, i.e., irreducible stochastic variability.
Despite this close conceptual alignment, the link between internal climate variability and aleatoric uncertainty has not yet been formalized.
We establish a theoretical link by showing that member-to-member differences in single-model large ensembles provide a direct representation of aleatoric uncertainty.
To quantify the spatio-temporal structure of aleatoric uncertainty, we employ generalized additive models.
The proposed framework is validated through comparison with ERA5-Land reanalysis data, demonstrating that ensemble-derived estimates reproduce key spatial and temporal patterns of real-world variability.
Applied to the water balance over the Iberian Peninsula, our approach reveals coherent variability structures and pronounced regional heterogeneity.
We find a decline in variability in drought-prone regions and seasons, a pattern that strengthens under +3\degree{C} global warming, implying an increased risk of persistent summer drought conditions.
Beyond this application, the framework is climate-model agnostic and transferable to other variables and spatial scales, providing a statistical basis for quantifying internal climate variability as aleatoric uncertainty.
      \end{abstract}
    \end{minipage}
  \end{center}

  \begin{center}
    \begin{minipage}{0.92\linewidth}
      \smallskip
      \noindent\textbf{Keywords:} generalized additive models, aleatoric uncertainty, natural variability, uncertainty quantification, spatio-temporal modelling
      \bigskip
    \end{minipage}
  \end{center}
}]   
\renewcommand\thefootnote{}
\footnotetext{\textbf{Abbreviations:}
CRCM5-LE, Canadian Regional Climate Model Large Ensemble; 
GAM, Generalized Additive Model; 
IPCC, Intergovernmental Panel on Climate Change; 
RCP8.5, Representative Concentration Pathway~8.5}

\renewcommand\thefootnote{\fnsymbol{footnote}}
\setcounter{footnote}{1}

\section{Introduction}\label{sec:intro}

Uncertainty plays a central role in climate science and is critical for informing robust and explainable policy decisions. As \citet{cripps2023UncertaintyNothingMore} emphasize, decision makers rely on a clear understanding of the uncertainties inherent in environmental information, particularly when long-term projections guide high-stakes planning, risk management, and adaptation \citep{lehner2023origin}.

A major source of uncertainty in climate trajectories is \emph{natural climate variability}---fluctuations in climate variables arising from natural processes intrinsic to the chaotic climate system \citep{feldstein2000,madden1976}. Natural variability sets a fundamental limit to predictability, rooted in the chaotic dynamics of the atmosphere \citep{lorenz1963DeterministicNonperiodicFlow, palmer2000PredictingUncertaintyForecasts}, and understanding it is essential for the robust detection of anthropogenic climate signals \citep{bindoff2013, santer2019quantifying}. Yet this detection is far from trivial: natural variability may obscure or amplify long-term climate signals. A useful framing is therefore to view the problem as one of signal versus noise, with natural variability representing the noise that must be disentangled from the climate-change signal \citep{hawkins2009, hawkins2011}.

\begin{figure*}[t!]
    \centering
    \includegraphics[width=0.99\linewidth, trim={0 35 40 60}, clip]{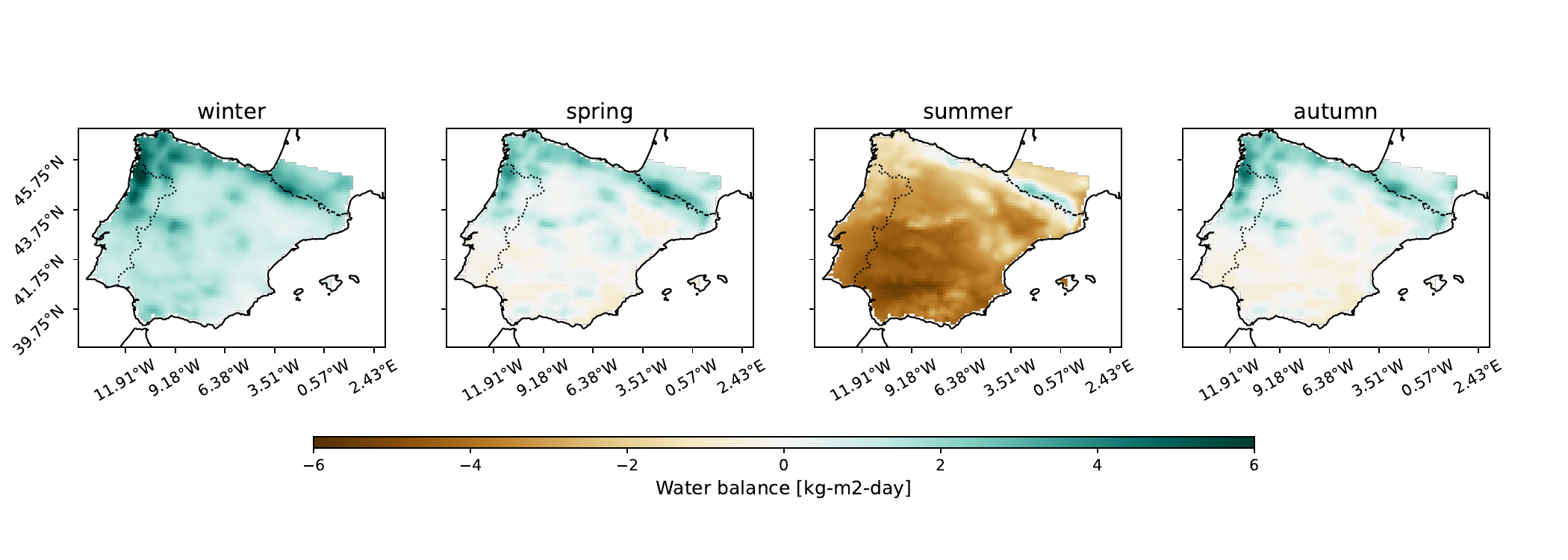}
    \caption{Climatological mean seasonal water balance over the Iberian Peninsula for 1991--2020, derived from ERA5-Land reanalysis. Positive values indicate moisture surplus, while negative values reflect moisture deficits.}
    \label{fig:waterbalance_ip}
\end{figure*}

Natural variability is commonly separated into internal and external components. 
Internal variability arises from unforced fluctuations driven by the inherently chaotic dynamics of the climate system, including large-scale modes such as the El Niño Southern Oscillation and interactions within the coupled ocean–atmosphere system \citep{deser2010seasonal}. 
External variability, in contrast, results from exogenous forcings such as volcanic eruptions or solar variability \citep{ipcc2023ClimateChange2021}.

A seminal contribution in this field is due to \citet{hawkins2009, hawkins2011}, who introduced a decomposition of climate projection uncertainty into contributions from internal variability, model uncertainty, and scenario uncertainty by estimating the externally forced signal using a temporal spline. Their additive framework has since been widely applied and extended \citep{lehner2023new}, for example, by incorporating uncertainties induced by bias correction \citep{wu2022} or by allowing for time-varying internal variability through rolling averages \citep{lafferty2023}.

Single-model initial-condition large ensembles have become an essential tool for quantifying internal climate variability. Using a 40-member ensemble \citep{gent2006}, \citet{deser2012} quantify responses to low-frequency variability through linear regression and mean differences across present (2005 to 2014) and future (2051 to 2060) periods. Their results reveal substantial regional discrepancies and indicate that internal variability can account for at least half of the model spread. A related line of work employs dynamical adjustment techniques, such as the regularized (non-)linear models of \citet{sippel2019}, to disentangle natural atmospheric circulation variability from externally forced responses in temperature and precipitation.

More broadly, \citet{deser2020InsightsEarthSystem} highlight the importance of distinguishing between uncertainties arising from natural variability and those due to differences in model formulation.
The statistics and machine learning literature provides a conceptual basis for this challenge through the lens of uncertainty quantification. Uncertainty can be (not necessarily additively) separated into epistemic uncertainty, which stems from incomplete knowledge of processes, and aleatoric uncertainty, which reflects the irreducible randomness in data \citep{hullermeier2021AleatoricEpistemicUncertainty, gruber2025SourcesUncertaintySuperviseda}. From this probabilistic perspective, natural variability can be interpreted as the realization of aleatoric uncertainty in the climate system \citep{singh2019QuantifyingUncertaintyTwentyfirst, verjans2025GreenlandIceSheet}.
However, this interpretation relies on the assumption that climate models adequately represent the underlying stochastic dynamics, an aspect we revisit in Section \ref{ssec:lim} when discussing the role of epistemic uncertainty.
This distinction has recently also been discussed in the context of weather and climate modeling \citep{mansfield2025EpistemicAleatoricUncertainty}.

Following \citeauthor{bonas2025AssessingPredictabilityEnvironmental}'s \citeyearpar{bonas2025AssessingPredictabilityEnvironmental} call to formalize and strengthen uncertainty quantification in environmental statistics, our study contributes to the climate-uncertainty literature in three ways: First, we present a rigorous statistical framework for extracting aleatoric uncertainty from climate model ensembles by separating the variability of a climate variable from its mean state. The framework is applicable across climate variables, spatial regions, and ensemble designs. Second, we introduce a validation strategy that uses reanalysis data to verify the separation of systematic bias from natural variability. Third, we provide a specific application example of this approach by conducting a detailed quantification of water-balance uncertainty across the Iberian Peninsula. This application example illustrates the spatial and temporal structure of natural variability in a drought-prone region and underlines the importance of separating epistemic (model uncertainty) and aleatoric (natural variability) uncertainty.

Together, these contributions establish a coherent basis for probabilistic assessments of climate uncertainty. As climate information increasingly informs adaptation planning and other high-stakes decisions, rigorous and transparent quantification of aleatoric uncertainty becomes, in our view, an essential component of responsible climate science.

\section{Data}\label{sec:data}

\subsection{Regional Domain and Climate Variable}
To demonstrate the utility of our framework, we apply it to the use case of water balance over the Iberian Peninsula, a region recognized as a hotspot of hydroclimatic vulnerability.

Water balance, defined as precipitation minus potential evapotranspiration derived from temperature \citep{thornthwaite1948}, is a fundamental indicator of water availability and drought risk \citep{xu1998ReviewMonthlyWater}. In Mediterranean climates, fluctuations in water balance have a critical influence on agricultural productivity, water management, and ecosystem stability \citep[e.g.][]{garrote2017managing, almendra2021agricultural}. Its sensitivity to both mean-state changes and internal variability makes it a particularly suitable testbed for disentangling aleatoric uncertainty.

We analyze monthly water-balance variability for two 30-year climate periods:
\begin{itemize}
    \item \textit{present}: current climate conditions under approximately 1.2\degree{C} global warming (1991--2020),
    \item \textit{future (3-degree)}: projected conditions under roughly 3\degree{C} global warming. For our model (see \Cref{sec:results_future}), this corresponds to the period: 2037--2066.
\end{itemize}
Using global warming levels enables a consistent comparison of climate-change impacts across periods \citep{seneviratne2021, gampe2024applying}.

The spatial domain corresponds to the PRUDENCE definition of the Iberian Peninsula \citep{christensen2007, bohnisch2021hot}. \autoref{fig:waterbalance_ip} illustrates the seasonal and regional climatology of water balance on the Iberian Peninsula for the years from 1991 until 2020. The pronounced spatial gradients and strong seasonal cycle underscore the region’s exposure to summer droughts, highlighting the necessity of rigorous uncertainty quantification for informing climate-impact assessments and adaptation planning \citep{iglesias2000AgriculturalImpactsClimate}.

\subsection{CRCM5 Large Ensemble}

\begin{figure}[htb]
    \centering
    \includegraphics[width=0.85\linewidth, trim=20mm 85mm 70mm 65mm, clip]{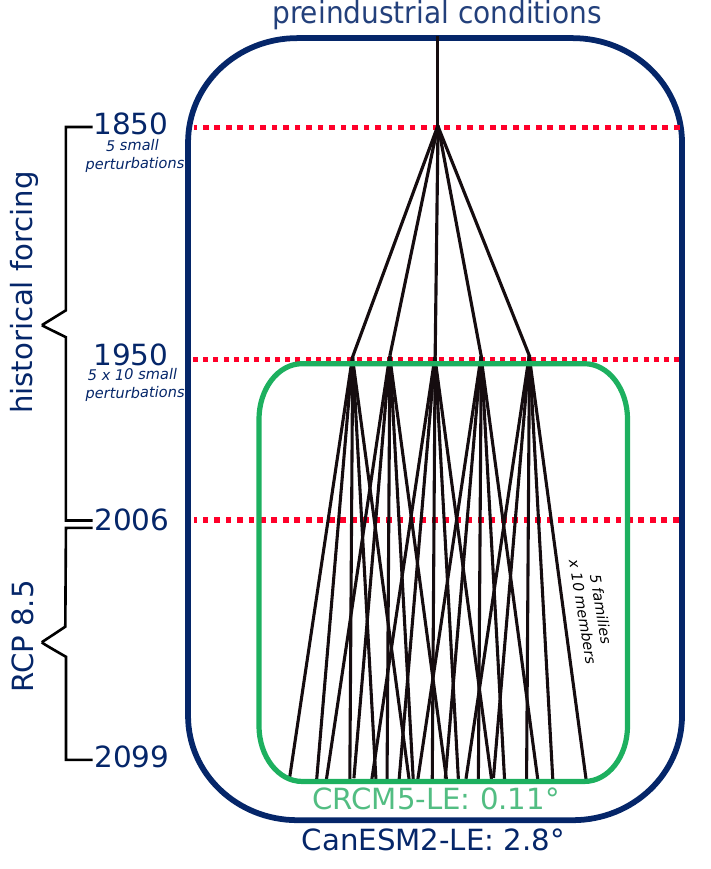}
    \caption{Conceptual schematic showing the organization of climate model families and ensemble members in the driving global large ensemble (CanESM2-LE, 2.8\degree{} resolution) and the dynamically downscaled regional large ensemble (CRCM5-LE, 0.11\degree{} resolution). Adapted from \cite{sasse2023low, poschlod2020impact}.}
    \label{fig:CRCM5_member_family}
\end{figure}

Single model initial-condition large ensembles have emerged as a powerful approach for separating uncertainty sources by running multiple, equally likely realizations of the same model with slightly perturbed initial conditions \citep{maher2021}. The spread across ensemble members is driven by identical external forcings and provides direct estimates of internal climate variability, enabling the quantification of aleatoric uncertainty \citep{deser2020InsightsEarthSystem,kay2015CommunityEarthSystem, thompson2015QuantifyingRoleInternal}.

In this study, we utilize the Canadian Regional Climate Model Large Ensemble (CRCM5-LE; \citealp{leduc2019ClimExProject50Member, martynov2013reanalysis, vseparovic2013present}), a single model initial-condition large ensemble designed for regional analyses of extreme events and internal climate variability \citep[e.g.][]{willkofer2020holistic, funk25}. The regional climate model CRCM5-LE dynamically downscales its driving global climate model, the Canadian Earth System Model version~2 (CanESM2), over the European domain to a fine spatial resolution of 0.11\degree{} (approximately 12\,km). 

The ensemble comprises 50 members organized into five families, each containing 10 perturbations. Each family is generated from small perturbations to the initial conditions of CanESM2 in 1850, while individual members within a family arise from additional perturbations applied in 1950 (see \autoref{fig:CRCM5_member_family}). All members share identical model physics and external forcing but diverge solely due to these perturbations, thereby sampling a large set of equally plausible climate trajectories. Member trajectories can therefore be assumed to be independent, equally likely realizations of the same chaotic climate system after a five year spin-off phase in the last perturbation in 1950 \citep{leduc2019ClimExProject50Member}.
We analyze data from 1991 onwards, for which family effects are negligible. 
The large ensemble size (50 members) enables robust estimation of climate trajectories and extremes with a total of 1500 modeled years for the quantification of a climatic period of 30 years.
From 2006 onward, external forcings follow the Representative Concentration Pathway~8.5 (RCP8.5) scenario developed by the Intergovernmental Panel on Climate Change (IPCC). Further details on the construction of CRCM5-LE and the derivation of its boundary conditions from CanESM2 are provided in \cite{fyfe2017large, leduc2019ClimExProject50Member}.

The CRCM5-LE has been extensively used to analyze water-balance-relevant climate variables such as temperature and precipitation \cite[e.g.][]{bohnisch2021hot, bohnisch2023european, felsche2024european}. Compared with global climate models, its higher spatial resolution enables a much more detailed representation of topographic and regional influences on these variables and other drought-related indices, which is an important advantage given the substantial elevation gradient in our study area, ranging from sea level to 2,731 m. Consequently, CRCM5-LE is well-suited for analyzing local hydrometeorological processes and extremes, including floods and droughts \citep{poschold2025, mittermeier2019detecting, willkofer2024assessing}.
The set of model trajectories provides a substantially larger data basis than observational or reanalysis data, enabling a statistically robust estimation of internal variability \citep{vontrentini2019assessing}. While this variability represents \emph{pseudo}-aleatoric uncertainty arising from model realizations, it offers a controlled framework for isolating internal climate variability. 

\subsection{ERA5-Land Reanalysis}
In \Cref{sec:val_section}, we validate our results against the monthly ERA5-Land (ERA5) reanalysis data, the fifth-generation reanalysis produced by the European Centre for Medium-Range Weather Forecasts \citep{hersbach2020ERA5GlobalReanalysis, munoz2021}.
ERA5-Land provides a comprehensive global dataset of historical climate and weather variables by integrating observational data with numerical weather models through data assimilation. This approach addresses the inherent spatial heterogeneity in observational networks, which exhibit significant gaps over sparsely populated areas. Thus, ERA5-Land is a spatially and temporally consistent and physically coherent representation of Earth's atmosphere and land surface, offering superior global coverage compared to raw observational data alone.
For the validation framework, we use ERA5-Land water balance data for the \textit{present} period (1991--2020). 
The ERA5-Land dataset is originally provided at a native spatial resolution of approximately 9\,km on a reduced Gaussian grid and is subsequently regridded to match the spatial resolution of the CRCM5-LE grid (~12.5 km, EUR-11 rotated-pole grid). Prior to interpolation, the ERA5-Land domain is transformed to the EUR-11 rotated-pole coordinates. The remapping is performed using bilinear interpolation \citep{schulzweida_2023}, with the CRCM5-LE domain grid serving as the target. This harmonization of resolution ensures spatial consistency and enables a robust comparison between the two datasets.

\section{Theoretical Framework}
\label{sec:theoretical_framework}

\subsection{Uncertainty Decomposition in Climate Ensembles}

Climate model output can be conceptually decomposed into systematic and stochastic components that reflect different sources of variability and uncertainty. To clarify our terminology, we refer to each individual simulation within the climate model ensemble as an ``ensemble member''. For a given climate variable $Y$ from ensemble member $i$, at location or space $s$, and time $t$, we propose the following decomposition:
\begin{equation} \label{eq:y_psi_theta_members}
    Y_{i}(s,t) = \psi(s,t) + \vartheta_{i}(s,t),
\end{equation}
where $Y_{i}(s,t)$ represents the observed climate variable (in our case, water balance), $\psi(s,t)$ denotes the systematic component that captures the deterministic model response to external forcings and boundary conditions, and $\vartheta_{i}(s,t)$ represents the stochastic component arising from internal climate variability.

The systematic component $\psi(s,t)$ encompasses all deterministic model behavior, including the mean state of the climate variable and systematic climate model biases, inducing epistemic uncertainty. The stochastic component $\vartheta_{i}(s,t)$ captures the model's interpretation of internal climate variability, causing what we term aleatoric uncertainty.
In this study, we aim to quantify aleatoric uncertainty in order to characterize internal climate variability.
To do so, our methodological approach relies on three key assumptions that enable the separation of systematic and stochastic components.

\textbf{Assumption 1:} The systematic component $\psi(s,t)$ is identical across all ensemble members, reflecting the fact that all members use identical model physics, parameterizations, and boundary conditions. This assumption is fundamental to large ensemble methodologies and is supported by the ensemble generation protocol, where members differ only in initial conditions. While $\psi(s,t)$ may, in principle, contain member-family-specific structure, such effects are irrelevant for our analysis. Specifically, we analyze data from 1991 onward, i.e., more than 30 years after the initialization of different ensemble families, a time horizon over which any residual dependence on the initial-condition family is expected to have vanished \citep{leduc2019ClimExProject50Member}. Moreover, even if a family-specific component persisted, it cancels out in the pairwise member differences considered in the next section. We therefore treat $\psi(s,t)$ as a nuisance component, as our primary interest lies not in potential biases or mean climate states, but in the variability $\vartheta_{i}(s,t)$.

\textbf{Assumption 2:} The stochastic components $\vartheta_{i}(s,t)$ and $\vartheta_{j}(s,t)$ for two arbitrary ensemble members $i$ and $j$ are independent realizations of the same underlying natural variability process. This independence assumption is justified by the chaotic nature of atmospheric dynamics, where small differences in initial conditions lead to divergent but equally plausible climate trajectories.

\textbf{Assumption 3:} The stochastic component $\vartheta_{i}(s,t)$ has zero mean, $E[\vartheta_{i}(s,t)] = 0$, and variance $\text{Var}[\vartheta_{i}(s,t)] = \sigma^2_{\vartheta}(s,t)$, which may vary in space and time but is consistent across ensemble members. This assumption reflects the symmetric nature of internal variability around the systematic model response. 
This assumption is considerably weaker than the normality assumptions typically adopted in the literature, for example in \cite{singh2019QuantifyingUncertaintyTwentyfirst, thompson2015QuantifyingRoleInternal}.

\subsection{Variance Extraction by Ensemble Differencing}

The core insight of our methodology lies in the mathematical properties of ensemble member differences. For two ensemble members $i$ and $j$ at a fixed space $s$ and time $t$, we consider the difference
\begin{equation}
    \delta_{i,j}(s,t) = Y_{i}(s,t) - Y_{j}(s,t).
\end{equation}
We define the pairwise differences as a spatio-temporal process 
$\{\delta_{i,j}(s,t): (s,t) \in \mathcal{S} \times \mathcal{T}\}$, with spatial domain $\mathcal{S}$ and time domain $\mathcal{T}$.
Using our decomposition of $Y_{i}(s,t)$, this becomes
\begin{align} \label{eq:var_ext}
    \delta_{i,j}(s,t) &= [\psi(s,t) + \vartheta_{i}(s,t)] - [\psi(s,t) + \vartheta_{j}(s,t)] \nonumber \\
    &= \vartheta_{i}(s,t) - \vartheta_{j}(s,t).
\end{align}

Thus, under the first assumption, the systematic component $\psi(s,t)$ cancels when computing differences between ensemble members, leaving only the difference between stochastic components. This property forms the theoretical foundation for our uncertainty quantification approach.

To quantify the magnitude of natural variability, we analyze the squared differences $\delta_{i,j}^2(s,t)$. Using assumptions 2 and 3, i.e., independence and zero-mean with identical variance,
\begin{align} \label{eq:nat_var_true}
    E[\delta_{i,j}^2(s,t)] &= E[(\vartheta_{i}(s,t) - \vartheta_{j}(s,t))^2] \nonumber\\
    &= E[\vartheta_{i}^2(s,t)] - 2E[\vartheta_{i}(s,t)\vartheta_{j}(s,t)] + E[\vartheta_{j}^2(s,t)] \nonumber\\
    &= \sigma^2_{\vartheta}(s,t) + \sigma^2_{\vartheta}(s,t) = 2\sigma^2_{\vartheta}(s,t),
    \end{align}
the expected value of the squared difference $E[\delta_{i,j}^2(s,t)]$ provides a direct estimator of the underlying internal variability.

This shows that the squared difference between ensemble members provides an unbiased estimator of twice the internal variability variance. Consequently, the standard deviation, capturing the aleatoric uncertainty, is:
\begin{equation} \label{eq:sigma_theoretical}
    {\sigma}_{\vartheta}(s,t) = \sqrt{\frac{1}{2} \; {E \left[\delta_{i,j}^2(s,t)\right]}}.
\end{equation}

\subsection{Generalized Additive modeling}

Generalized Additive Models \citep[GAMs;][]{hastie1986GeneralizedAdditiveModels,wood2017GeneralizedAdditiveModels} arise as a natural extension of classical regression models. 
While linear models assume a Gaussian response and linear effects of covariates, generalized linear models allow for a wider class of response distributions through a link function. 
GAMs further generalize this framework by replacing linear effects with smooth functions of covariates, enabling flexible, data-driven modeling of non-linear relationships.
We employ GAMs to model the expected value of squared differences as a smooth function of spatio-temporal covariates.
We analyze squared differences between two ensemble members, where $i$ and $j$ form a disjunct pair within the same family (see \autoref{fig:CRCM5_member_family} for the member-family structure and \autoref{tab:member_pairs} for the definition of all member pairs).
Our GAM specification assumes that squared differences follow a Gamma distribution with log link function:
\begin{align}
    \delta_{i,j}^2(s,t) \sim& \text{Gamma}(\mu_{i,j}(s,t), \phi) \\
    \log(\mu_{i,j}(s,t)) =& \beta_0 + \beta_1 \cdot \text{year} + f_1(\text{month}) + f_2(\text{elevation}) \nonumber \\ 
    &+  \mathbbm{1}_\text{season} \cdot f_{\text{season}}(\text{lon}, \text{lat}) + u_{i,j} \label{eq:gam_members},
\end{align}
where $\mu_{i,j}(s,t) = E [\delta_{i,j}^2(s,t)]$, 
with a linear temporal trend with slope $\beta_1$, a smooth cyclical effect for month $f_1$, a smooth effect for topology $f_2$, a tensor product $f_{\text{season}}$ for spatio-temporal effects, consisting of latitude ($lat$) and longitude ($lon$) per season and a random intercept $u_{i,j}$ accounting for unspecified systematic differences between member pairs \citep{wood2006LowRankScaleInvariantTensor}.

The Gamma distribution is appropriate for modeling positive-valued squared differences, while the log link ensures positive predictions and enables multiplicative interpretation of covariate effects. The specific functional forms capture hypothesized sources of uncertainty variation, including spatial patterns that vary by season, cyclic monthly effects, elevation dependencies reflecting topographic influences, linear temporal trends potentially related to climate change, and member-pair specific effects. 

Finally, we are interested in the internal variability, or aleatoric uncertainty,  given in Eq.~(\ref{eq:nat_var_true}) and estimate it by:
\begin{align}
     \hat{\sigma}_{\vartheta}(s,t) = \sqrt{\frac{1}{2} \; {\hat{\mu}_{i,j}(s,t)}}.
\end{align}

Before turning to the results, we briefly comment on the spatio-temporal dependence structure implied by our model.
Rather than explicitly specifying a covariance function $\mathrm{Cov}\big(\delta_{i,j}(s,t), \delta_{i,j}(s',t')\big)$ as in classical geostatistical approaches, we capture spatial and temporal dependence implicitly through two complementary mechanisms.
This is motivated by the duality between mean and covariance modeling noted by \citet{cressie2015statistics}: ``what is one person's (spatial) covariance structure may be another person's mean structure''.
The B-spline functions of spatial and temporal covariates in model~(\ref{eq:gam_members}) encode autocorrelation structure directly into the mean through their smooth mean specification \citep{Wang15032024}.
Additionally, in GAM estimation, penalized spline terms can be interpreted as random effects, thereby inducing an implicit correlation structure across space and time \citep{wood2011fast}.
Together, these mechanisms account for spatio-temporal dependence without requiring an explicit covariance model.

\section{Results: Spatio-temporal Patterns of Internal Climate Variability}
\label{sec:results_present}

In the following, we present results from model Eq.~(\ref{eq:gam_members}) fitted to present-day CRCM5-LE data. Validation of these results is provided in \Cref{sec:val_section}. 
Model estimation is carried out using generalized additive models as implemented in the \texttt{mgcv} package in R \citep{wood2015GeneralizedAdditiveModels}. In this part, the data comprise 30 years $\times$ 12 months, 71 latitude pixels, and 96 longitude pixels, which result in 4,136 unique spatial grid points. For 25 ensemble member pairs, this yields a total of 37,224,000 observations. Given the large scale of the dataset, efficient estimation is essential; we therefore rely on computationally scalable fitting procedures tailored to high-dimensional data. Details on the estimation algorithm and implementation are provided in Appendix~\ref{sec:appendix_model_fitting}.

\paragraph*{Goodness of Fit}

The generalized additive model specified in Equation~\ref{eq:gam_members} explains 27.3\% of the deviance in squared ensemble member differences. While this might appear modest, it is important to recognize that internal variability is inherently dominated by unstructured stochastic fluctuations. Our model deliberately focuses on capturing systematic patterns that vary coherently across space and time, thereby isolating the predictable components of aleatoric uncertainty from pure noise. The remaining unexplained deviance represents the irreducible random component that lacks spatial or temporal structure.

\paragraph*{Temporal Trends}

We first analyze the magnitude and the long-term trend of internal variability as represented in model (\ref{eq:gam_members}). This represents the base level of estimated variability, computed as:
\begin{align}
     \hat{\sigma}_{\vartheta}(\text{year}) = \sqrt{\frac{1}{2} \; \exp\left(\hat{\beta}_0 + \hat{\beta}_1  \cdot \text{year}\right)}.
\end{align}
 
Across the 30-year period, the inferred internal variability of water balance in the Iberian Peninsula remains essentially constant, increasing only slightly from 1.585 kg m$^{-2}$ day$^{-1}$ in 1991 to 1.589 kg m$^{-2}$ day$^{-1}$ in 2020. This serves as the baseline against which the seasonal, topographic, and spatial patterns discussed later are compared.

\paragraph*{Seasonality of Variability}

\begin{figure}[t]
    \centering
    \includegraphics[width=1\linewidth]{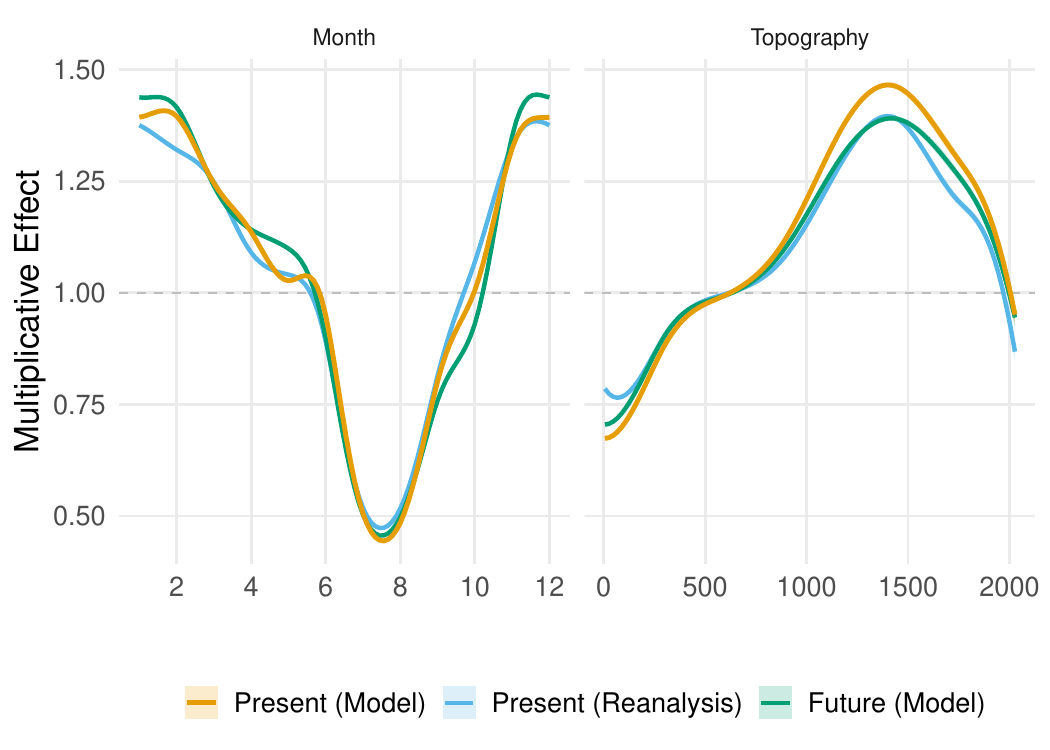}
    \caption{Comparison of the estimated multiplicative effects of month and topography on the base magnitude of internal variability. The figure contrasts effects derived from ensemble members in the \textit{present} climate model world, from model-reanalysis comparisons in the \textit{present}, and from climate model ensemble members under a \textit{future (3-degree)} global warming level. The curves highlight consistent seasonal and topographic patterns across members while revealing systematic similarities and differences between \textit{present}, \textit{3-degree} warming and model-reanalysis. Note that the topographic effect represents the isolated influence of elevation on internal variability, independent of geographical location.}
    \label{fig:comparison_splines}
\end{figure}

The estimated multiplicative monthly effect, 
$$\sqrt{\exp\left(\hat{f_1}(\text{month}\right)},$$ 
derived from Eq.~(\ref{eq:gam_members}), is shown in \autoref{fig:comparison_splines} (left panel, orange line). The remaining curves shown in the figure are discussed in detail in \Cref{sec:results_validation} and \Cref{sec:results_future}. The estimate reveals a pronounced seasonal cycle in internal variability. Summer months, particularly July and August, exhibit the lowest variability with multiplicative effects of 0.48 and 0.51, respectively, representing approximately a 50\% reduction relative to the annual mean. This attenuation corresponds to the dry season in Mediterranean climates, during which water balance variability is constrained by the near-absence of precipitation events. As a result, our analysis indicates that summers on the Iberian Peninsula are consistently dry, with little inter-annual variation in water balance conditions (see also \autoref{fig:waterbalance_ip}).

In contrast, variability peaks during late autumn and winter, with November through February showing the highest multiplicative effects, ranging from 1.32 to 1.40. These months exhibit 32\% to 40\% greater variability than the annual mean, reflecting the period of highest precipitation sums in the Iberian Peninsula, and when Atlantic disturbances are most active and thus precipitation patterns are most uncertain \citep{serrano1999MonthlyModesVariation, trigo2001PrecipitationScenariosIberia}. Spring months display a gradual decline in variability, with March showing moderately elevated effects (1.25) that decrease through April (1.14) to near-average levels by May (1.03) and June (0.95). The autumn transition months of September and October show intermediate variability (0.80 and 1.01, respectively), marking the shift from the stable summer regime to the more dynamic winter period. 

\paragraph*{Topographic Influences}

In the GAM framework, we separate the spatial structure into a latitude--longitude effect and a separate one-dimensional elevation smooth, with the latter capturing effects driven solely by elevation and independent of geographic location.
The estimated multiplicative elevation effect, 
$$\sqrt{\exp\left(\hat{f_2}(\text{elevation}\right)},$$ 
derived from Eq.~(\ref{eq:gam_members}), exhibits a pronounced nonlinear relationship with internal variability (\autoref{fig:comparison_splines}, right panel, orange line). At low elevations near sea level, variability is reduced with a multiplicative effect of 0.67. The effect increases monotonically through mid-elevations, crossing unity around 600 meters and reaching a maximum of 1.47 at approximately 1,400 meters above sea level. Beyond this peak, variability gradually declines at higher elevations, returning to near-baseline levels (0.95) at 2,000 meters.

This pattern suggests that mid-elevation mountainous terrain exhibits the greatest variability in water balance, likely due to the complex nature of orographic precipitation and evapotranspiration \citep{roe2005OROGRAPHICPRECIPITATION}. The reduced variability at sea level may reflect more homogeneous atmospheric conditions, while the decline at the highest elevations suggests a transition towards more consistent high alpine regimes with low variability in precipitation \citep{lopez2008}. 

\paragraph*{Spatial Heterogeneity in Natural Variability}
\begin{figure*}[!ht]
    \centering
    \includegraphics[width=\linewidth]{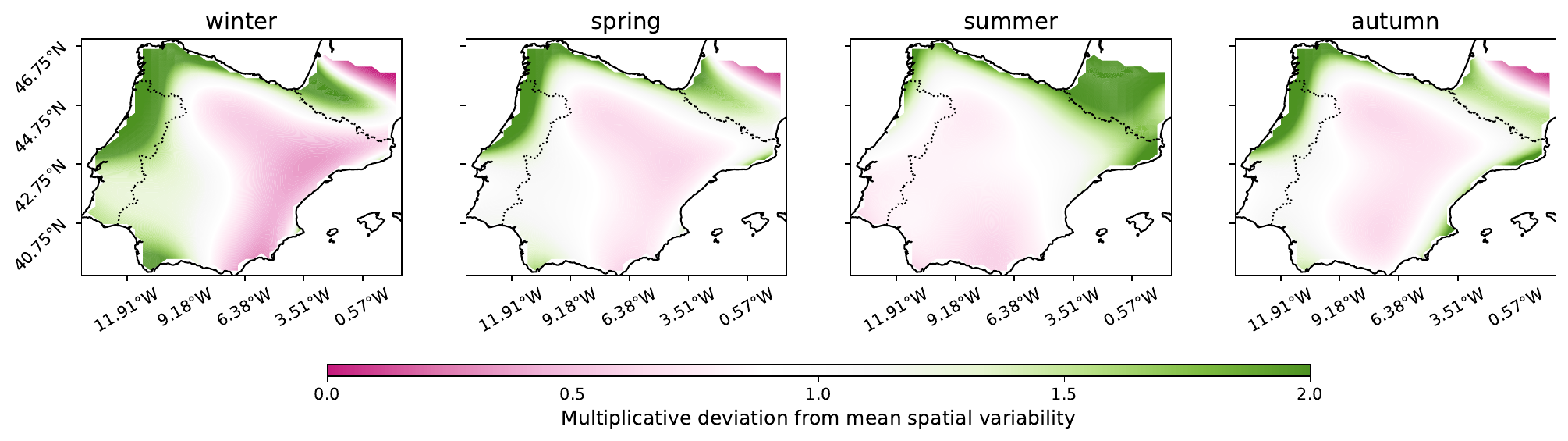}
    \includegraphics[width=\linewidth]{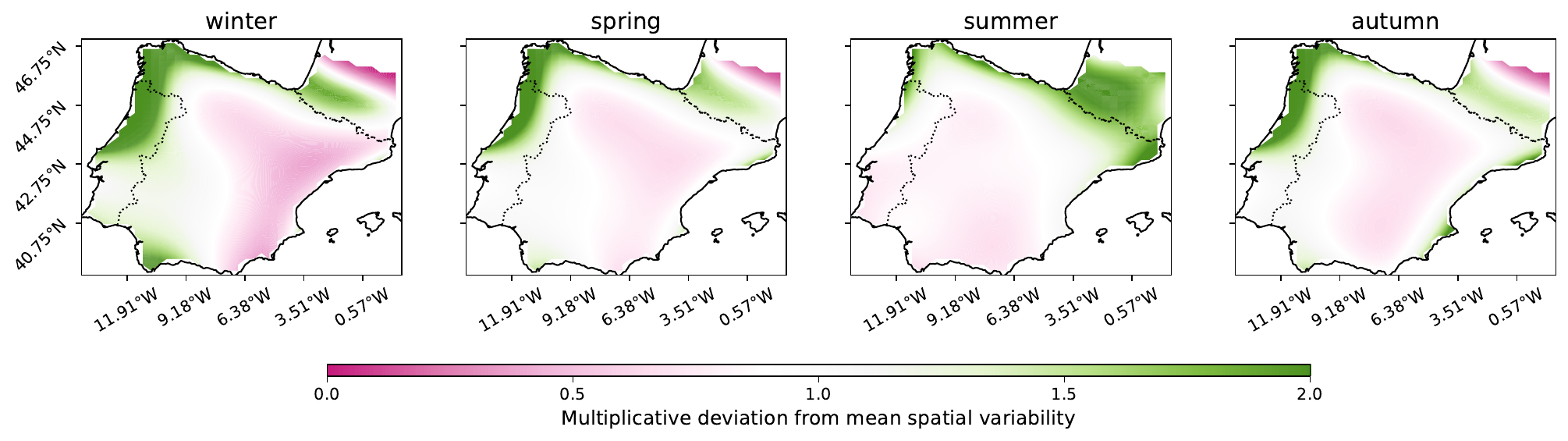}
    \includegraphics[width=\linewidth]{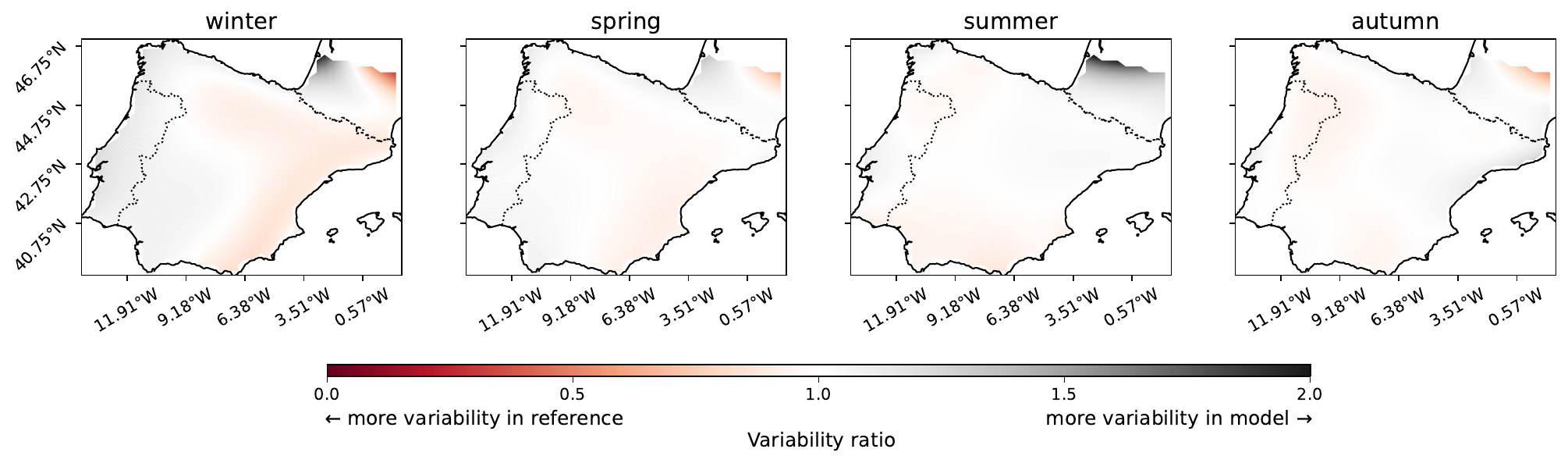}
    \caption{Spatial distribution of the estimated multiplicative effects on internal variability across seasons. The top row shows the multiplicative deviation from mean spatial variability derived from ensemble members in the \textit{present} climate model world. The middle row displays corresponding deviations obtained from model-reanalysis comparisons in the \textit{present}. The bottom row presents the spatial ratio between the two, indicating where variability is higher in the reference (<1) or higher in the model (>1). Together, the panels demonstrate seasonal coherence in spatial patterns of climate variability while revealing where model-derived and reanalysis-derived estimates align or diverge across the Iberian Peninsula.}
    \label{fig:effects_members}
\end{figure*}

The estimated multiplicative spatial effect, 
$$\sqrt{\exp\left(\hat{f}_{\text{season}}(\text{lon}, \text{lat})\right)},$$ 
derived from Eq.~(\ref{eq:gam_members}) captures how variability is distributed across the Iberian Peninsula per season, included as a categorical variable dividing the year into four periods: winter (Dec--Feb), spring (Mar--May), summer (Jun--Aug), and autumn (Sept--Nov). The top panel of \autoref{fig:effects_members} displays these four seasonal spatial patterns; the middle and bottom panels are discussed in \Cref{sec:results_validation}.

The spatial maps reveal pronounced heterogeneity in internal variability across the Iberian Peninsula, with coastal regions exhibiting substantially higher uncertainty than the interior. The multiplicative spatial effects indicate that the northern coast experiences roughly twice the baseline variability year-round. A similar pattern emerges along the western coastline and the southern tip near the Strait of Gibraltar, where winter, spring, and autumn exhibit a twofold increase, although this amplification is notably absent in summer. In winter, these areas form a continuous band of elevated variability stretching from the northern coast down along the Portuguese shoreline to the southern coast, including the Gibraltar region.

This coastal amplification of variability likely reflects the influence of Atlantic weather systems, which introduce greater stochasticity in precipitation patterns near oceanic boundaries. In contrast, the southern and central interior regions display below-average variability, with multiplicative effects ranging from 0.6 to 0.8. While these regions face greater mean drought stress, the reduced variability indicates more predictable deviation patterns from climatological norms.

Overall, while the model captures the magnitude of variability and reveals coherent spatio-temporal patterns, the large remaining unexplained component indicates that internal climate variability is predominantly unstructured, providing quantitative evidence for its fundamentally aleatoric nature.

\section{Validation Using Reanalysis}
\label{sec:val_section}
\subsection{Theoretical Framework}
\label{sec:val}

To evaluate whether ensemble-difference-based variability estimates truly isolate internal climate variability, we validate our approach using ERA5-Land reanalysis. ERA5 provides independent, historical climate conditions based on the assimilation of observational data, whose structural errors differ fundamentally from those in CRCM5-LE. This independence makes ERA5 a suitable benchmark for testing whether ensemble differencing effectively removes systematic model components.

Analogous to Eq.~(\ref{eq:y_psi_theta_members}), the ERA5 water balance is decomposed into a systematic and a stochastic term,
\begin{equation} \label{eq:y_psi_theta_renalysis}
    \tilde{Y}(s,t) = \tilde{\psi}(s,t) + \tilde{\vartheta}(s,t),
\end{equation}
where $\tilde{\psi}$ denotes the climatological mean state including potential biases and $\tilde{\vartheta}$ represents natural variability.

Our validation mirrors the ensemble-based procedure but replaces one ensemble member with ERA5. The goal is to quantify (i) systematic differences between model and reanalysis and (ii) the residual variability after removing those differences. Agreement between the spatial and temporal patterns of this residual variability and those obtained from ensemble differencing would support the theoretical assumption that model member differences isolate internal variability.

\paragraph*{Step 1: Modeling Systematic Differences}

For each CRCM5-LE member $j$, we analyze the signed difference between ERA5 and the model,
\begin{align}
    \tilde{\delta}_{j}(s,t) &= \tilde{Y}(s,t) - Y_{j}(s,t) \nonumber \\
    & = \tilde{\psi}(s,t) - \psi(s,t) + \tilde{\vartheta}(s,t) -\vartheta_{j}(s,t).
\end{align}
which no longer cancels the systematic components. We therefore model $\tilde{\delta}_{j}(s,t)$ using a GAM capturing temporally, spatially, and topographically structured biases:
\begin{equation}
\label{eq:gam_bias_reanalysis}
    \begin{split}
    \tilde{\delta}_{j}(s,t) =&~ \alpha_0 + \alpha_1 \cdot \text{year} 
    + g_1(\text{month}) + g_2(\text{elevation}) \\
    &+ \mathbbm{1}_{\text{season}} \cdot g_{\text{season}}(\text{lon}, \text{lat}) + u_j 
    + \tilde{\varepsilon}_{j}(s,t),
    \end{split}
\end{equation}
with member-specific random intercepts $u_j \sim \mathcal{N}(0,\sigma_u^2)$.  
The residual term
\[
\tilde{\varepsilon}_{j}(s,t) = \tilde{\vartheta}(s,t) - \vartheta_j(s,t)
\]
represents the remaining stochastic discrepancy after systematic differences have been removed.

To ensure comparability with the ensemble-difference analysis, we use 25 members (five from each of the five CRCM5-LE families; see \autoref{tab:validation}). This matches the effective sample size obtained by pairing 50 ensemble members into 25 differences, ensuring comparable estimation uncertainty and statistical precision. Using all 50 members would increase precision, but at the cost of increased computational complexity and loss of direct comparability.

\paragraph*{Step 2: Quantifying Residual Variability}

After subtracting the estimated bias structure, the expectation of the squared residuals is
\begin{align}
    E[\tilde{\varepsilon}^2_{j}(s,t)] &= E[\tilde{\vartheta}(s,t) - \vartheta_{j}(s,t)]^2 \nonumber \\
     &= E[\tilde{\vartheta}^2(s,t)] - 2 E [\tilde{\vartheta}(s,t) \vartheta_{j}(s,t)] + E[ \vartheta_{j}^2(s,t)]\nonumber \\
     &= E[\tilde{\vartheta}^2(s,t)] + E[ \vartheta_{j}^2(s,t)] \nonumber \\
     &= \sigma^2_{\tilde{\vartheta}}(s,t) + \sigma^2_{\vartheta}(s,t),
\end{align}
using that the stochastic terms have zero mean.
If reanalysis-based natural variability coincides structurally and in magnitude with CRCM5-LE internal variability, i.e.,
$$
E \left[\delta_{i,j}^2(s,t)\right] \approx E\left[\tilde{\varepsilon}^2_{j}(s,t)\right],
$$
then the internal variability in ensemble members can be used as a proxy for natural climate variability:
$$
\sigma^2_{\tilde{\vartheta}}(s,t)\approx  \sigma^2_{\vartheta}(s,t).
$$
This is validated in the following, where
the squared residuals $\tilde{\varepsilon}^2_{j}(s,t)$ are assumed to follow a Gamma distribution, with $\tilde{\mu}_{j}(s,t) = E[\tilde{\varepsilon}^2_{j}(s,t)]$:
\begin{equation}\label{eq:valgamma}
    \tilde{\varepsilon}^2_{j}(s,t) \sim \text{Gamma}(\tilde{\mu}_{j}(s,t), \tilde{\phi}),
\end{equation}
with a log-link function modeled as:
\begin{align}
    \label{eq:gam_variability_reanalysis}
    \log(\tilde{\mu}_{j}(s,t)) =&~ \tilde{\beta}_0 + \tilde{\beta}_1 \cdot \text{year} 
    + \tilde{f}_1(\text{month}) + \tilde{f}_2(\text{elevation}) \nonumber \\
    &+ \mathbbm{1}_\text{season} \cdot \tilde{f}_{\text{season}}(\text{lon}, \text{lat}) + \tilde{u}_j,
\end{align}
where $\tilde{u}_j \sim \mathcal{N}(0, \tilde{\sigma}_u^2)$ allows for member-level heterogeneity across the ERA5 to CRCM5-LE differences.

This second-stage model mirrors the ensemble-based uncertainty analysis, enabling direct comparison of the estimated linear coefficients and smooth components, i.e., $\beta_k, f_k$ and $\tilde{\beta}_k, \tilde{f}_k$, respectively.
The comparison evaluates whether the spatial and temporal patterns of residual variability in the reanalysis-based model (representing \emph{real} aleatoric uncertainty) align with those from the ensemble-based model (representing \emph{pseudo}-aleatoric uncertainty).  
Agreement between the two validates the assumption that ensemble member differences faithfully extract internal variability.

\subsection{Results of Validation}
\label{sec:results_validation}

To assess whether our ensemble-based uncertainty estimates represent genuine internal variability rather than methodological artifacts, we compare the spatial and temporal patterns derived from ensemble member differences with those obtained from model-reanalysis discrepancies after bias correction, according to \Cref{sec:val}.

The temporal trend in overall variability magnitude, $\sqrt{\frac{1}{2}exp(\hat{\tilde{\beta}}_0 + \hat{\tilde{\beta}}_1 \cdot \text{year}}),$ 
in the ERA5-based estimate shows a modest declining trend, decreasing from 1.527 kg m$^{-2}$ day$^{-1}$ in 1991 to 1.465 kg m$^{-2}$ day$^{-1}$ in 2020, representing a 4\% reduction in the 30-year period. 
While the standard deviations estimated in the model and reanalysis deviate slightly (starting at 1.585 vs. 1.527 kg m$^{-2}$ day$^{-1}$, respectively), the magnitude is matched.
This minor divergence suggests that the reanalysis data indicate a slight decrease in observed natural variability over recent decades, whereas the climate model ensemble does not capture this trend. 

\autoref{fig:comparison_splines} (blue line) demonstrates that the monthly and elevation effects derived from $ \hat{\tilde{f}}_1(\text{month}) \text{ and } \hat{\tilde{f}}_2(\text{elevation})$ in model (Eq. \ref{eq:gam_variability_reanalysis}) align closely with those from ensemble member differences. The seasonal cycle shows the same summer minimum and autumn/winter maximum, with multiplicative effects differing by less than 0.1 across all months. The elevation effect similarly exhibits the monotonic increase peaking near 1,400 meters, though the ERA5-based estimates suggest a slightly less pronounced effect at elevations above 600 m.

The middle panel of \autoref{fig:effects_members} displays the multiplicative spatial effects estimated from  $\hat{\tilde{f}}_{\text{season}}(\text{lon}, \text{lat})$ in Eq. (\ref{eq:gam_variability_reanalysis}). The overall spatial structure closely resembles that derived from the CRCM5-LE model world (top panel), with coastal regions again showing elevated variability and interior regions displaying reduced variability. Thus, providing strong evidence that the ensemble-based approach successfully captures real-world spatial variability patterns across the Iberian Peninsula.

However, the bottom panel of \autoref{fig:effects_members}, which shows the ratio of ensemble-based to reanalysis-based variability estimates, reveals some systematic discrepancies. The model ensemble tends to slightly overestimate variability along portions of the northern and western coasts in winter, with ratios above 1.0, while showing good agreement in interior regions. These differences may reflect either limitations in CRCM5-LE's representation of coastal processes or uncertainties in ERA5's treatment of these transition zones.

Because the analysis relies on more than 37 million observations, we refrain from using formal significance tests for validation. With a sample size this large, even negligible differences between ensemble-based and reanalysis-based estimates would become statistically significant \citep{wasserstein2016ASAStatementPValues}, so we instead evaluate the magnitude and structural agreement of the estimated temporal, topographic, and spatial patterns.
Despite minor spatio-temporal deviations, there is an overall high correspondence between ensemble-based and reanalysis-based uncertainty patterns. This validates the theoretical framework's assumption that systematic biases cancel when computing the differences between ensemble members and that analyzing those differences directly yields uncertainty quantification of natural variability.

\section{Discussion}

\subsection{Scope and Limitations of the Framework}\label{ssec:lim}

Our theoretical framework provides a rigorous quantification of aleatoric uncertainty within the CRCM5-LE modeling system. However, several important limitations must be acknowledged. First, our results are strictly conditional on CRCM5's structural assumptions about climate physics and parameterizations. 
In particular, the interpretation of ensemble-member variability as aleatoric uncertainty relies on the assumption that the model adequately captures the stochastic, chaotic dynamics underlying natural variability. If this representation is misspecified, ensemble-derived variability may deviate from true natural variability and thus reflect, in part, epistemic uncertainty in the model formulation rather than purely aleatoric uncertainty \citep{jimenez2025WhyMachineLearning}.

To address this concern, we implement a validation strategy based on ERA5-Land reanalysis data, which provides an observationally constrained estimate of real-world variability. The close agreement in spatial, seasonal, and topographic patterns between ensemble-based and reanalysis-based estimates suggests that the model captures key structural features of natural variability. While this empirical validation cannot fully rule out misspecification of the underlying stochastic processes, it provides evidence that epistemic distortions in variability are limited and that ensemble differences offer a meaningful approximation of natural variability.

Furthermore, the additive decomposition in Equation (\ref{eq:gam_members}) assumes that systematic components and internal variability contribute additively, which may not hold for all variables or in all regions.

Despite these limitations, the proposed framework establishes a principled statistical approach to quantifying aleatoric uncertainty from climate model ensembles. It provides a flexible template that can be extended beyond the present application. In particular, applying the framework to multiple climate models would allow for a more explicit characterization of epistemic uncertainty arising from differences in model formulation. Similarly, extending the analysis to additional variables, impact-relevant indices, temporal resolutions, regions, and shared socioeconomic pathways would help assess the generality of the identified variability structures. Such extensions represent promising directions for future work and a natural next step toward a comprehensive decomposition of uncertainty in climate projections.

\subsection{Extension to Future Periods}
\label{sec:results_future}

\begin{figure*}[!ht]
    \centering
    \includegraphics[width=\linewidth]{figures/seasonal_variability_member.pdf}
    \includegraphics[width=\linewidth]{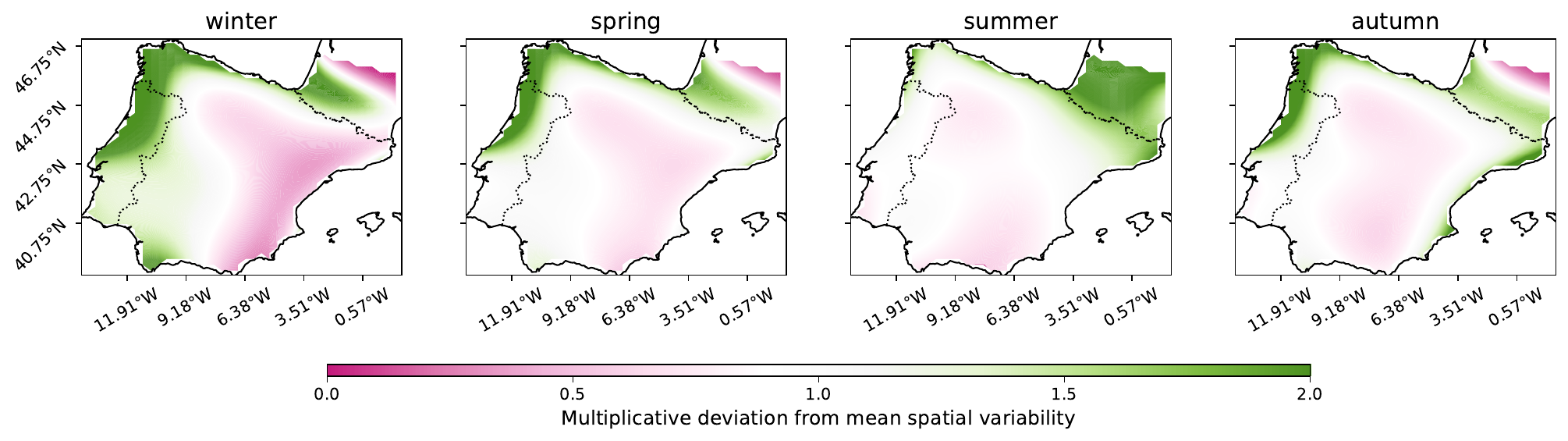}
    \includegraphics[width=\linewidth]{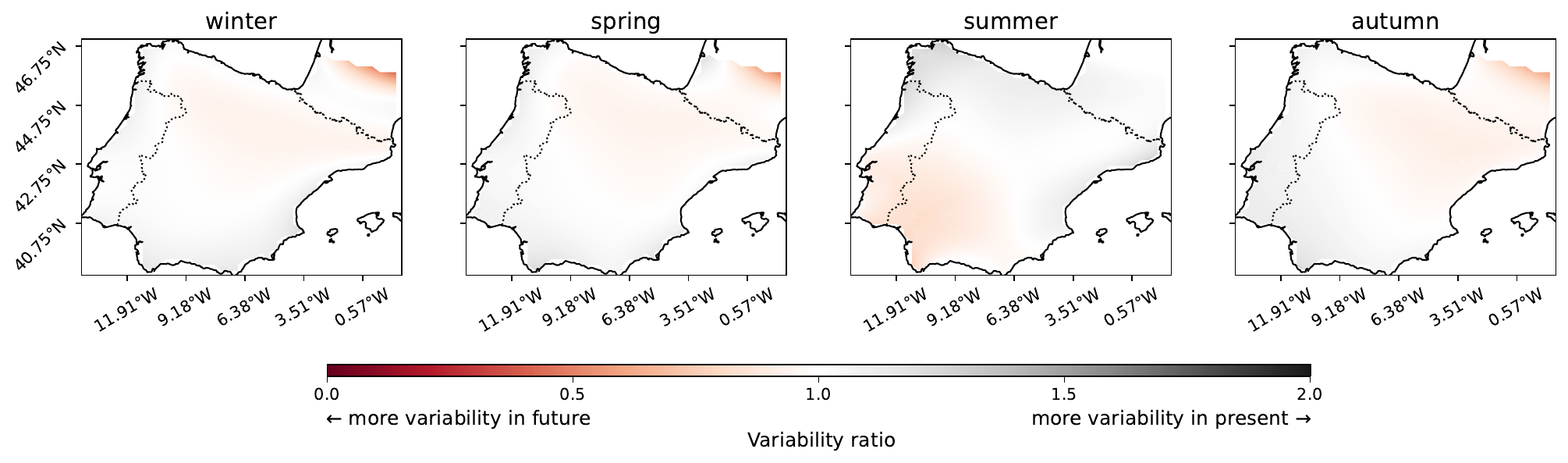}
    \caption{Spatial distribution of the estimated multiplicative effects on internal variability across seasons.
The top row shows the multiplicative deviation from mean spatial variability derived from ensemble members in the \textit{present} climate model world. The middle row displays corresponding deviations obtained from ensemble members under a \textit{3-degree} global warming level. The bottom row presents the spatial ratio between the two, indicating where variability is higher in the future warming level (<1) or higher in the present climate model world (>1). Together, the panels demonstrate seasonal coherence in spatial patterns of climate variability while revealing where present and future ensemble estimates align or diverge across the Iberian Peninsula.}
    \label{fig:effects_member_future}
\end{figure*}

Having established that the models reliably capture internal climate variability in the \textit{present} period, we now extend the analysis to \textit{future} projections. In the future, validation against observed climate is inherently impossible. The consistency demonstrated in the \textit{present} validation (\Cref{sec:val_section}) provides the basis for trusting that the modeled representation of internal variability remains meaningful under future climate conditions.

Applying the same methodology from \Cref{sec:theoretical_framework} to the future period 2037--2066, which corresponds to a \textit{3-degree} warming scenario, reveals notable shifts in both the magnitude and spatial distribution of internal variability between the \textit{present} conditions under +1.2\degree C and the future scenario under +3\degree C of global warming.

\paragraph*{Temporal Evolution of Uncertainty Magnitude}

The average water balance variability in the future period decreases from 1.521\,kg\,m$^{-2}$\,day$^{-1}$ in 2037 to 1.478\,kg\,m$^{-2}$\,day$^{-1}$ in 2066, corresponding to an approximate 3\% reduction over 30 years. A comparable decline of about 4\% is observed in the ERA5 reanalysis over the historical period from 1991 to 2020, whereas the ensemble-based estimate for the same period remains nearly constant. Despite differences in the present-period estimates, both reanalysis data and future projections consistently indicate a declining trend in the magnitude of variability. This agreement suggests that the reduction in variability may reflect a genuine response of the climate system, although the lack of a corresponding trend in the ensemble-based estimate introduces some uncertainty regarding the robustness of this signal.

\paragraph*{Seasonal Evolution}

The seasonal patterns of variability remain remarkably stable between the present and future periods (\autoref{fig:comparison_splines}, green line). The only notable difference occurs in May, where the future period shows a slightly higher multiplicative effect compared to the present, though this difference remains modest. This stability in the seasonal structure of natural variability suggests that the fundamental atmospheric dynamics governing intra-annual variability patterns are largely preserved under the projected future.

\paragraph*{Changes in Spatial Patterns}

The middle panel of \autoref{fig:effects_member_future} shows the spatial effects estimated for the future period. While the overall pattern of elevated coastal variability persists, the bottom panel reveals some regional differences when comparing future to present conditions. The ratio map indicates increased future variability (values below 1.0, shown in red) across the northern areas in winter, spring, and autumn. While in summer, the southern areas exhibit greater uncertainty.
This amplification may reflect an increasing frequency of extreme precipitation events punctuating longer dry periods, thereby widening the distribution of possible outcomes.

Conversely, in summer, areas along the northern margin exhibit decreased variability in the future period (values above 1.0, shown in black). This reduction in variability coincides with projections of decreased mean precipitation, suggesting a shift toward more persistent drought conditions that paradoxically reduce the range of natural variability around this drier mean state.

The spatial heterogeneity in how variability changes from present to future (\autoref{fig:effects_member_future}, bottom panel) underscores the importance of spatially resolved uncertainty quantification. Some regions experience increases in variability, while others show decreases. This heterogeneity implies that uniform uncertainty assumptions across the domain would substantially misrepresent the true state of knowledge about future water availability in specific locations.

\section{Conclusion}

This study establishes a rigorous statistical framework for quantifying aleatoric uncertainty in single model initial-condition large ensembles by separating irreducible internal variability from systematic model characteristics. Our approach leverages ensemble member differences to eliminate model mean state and bias, yielding direct estimates of internal variability through a generalized additive modeling framework that captures complex spatio-temporal patterns.

Applied to water balance over the Iberian Peninsula, the methodology reveals substantial heterogeneity in natural variability: coastal regions exhibit twice the baseline uncertainty of interior areas, and winter months show 40\% greater variability than the annual mean. 

Two-stage validation against ERA5-Land confirms the robustness of our approach and shows that ensemble-derived estimates reliably capture real-world internal variability, supporting our theoretical assumptions. While minor discrepancies remain in the overall magnitude and long-term trend of variability, the spatial patterns, seasonal cycle, and topographic gradients are reproduced with striking consistency between CRCM5-LE and the reanalysis data.

Projections under +3\degree C global warming reveal regionally divergent changes, with some areas experiencing variability increases exceeding 50\% while others show comparable decreases. This spatial heterogeneity has critical implications for climate adaptation: uniform uncertainty assumptions fundamentally misrepresent the true state of knowledge, and decision makers require spatially explicit quantification to appropriately weight climate information in high-stakes planning contexts.

The framework's model-agnostic design ensures broad applicability across climate variables, spatial domains, and climate ensemble architectures, extending naturally to multi-model intercomparison projects where distinguishing aleatoric from epistemic uncertainty becomes essential. The decomposition is applicable if at least two independent members share the same model bias and mean structure $\vartheta(s,t)$ (Equation~\ref{eq:var_ext}). Future applications could explore additional ensemble types, subdaily resolutions,  integration with dynamical adjustment techniques, or exploit additional information of different shared socioeconomic pathways. In particular, convection-permitting models could provide more detailed local evidence on extreme precipitation and its variability \citep{fosser2024, prein2015}.

As climate projections increasingly inform adaptation investments and resource management, transparent uncertainty quantification transitions from methodological refinement to ethical imperative. By providing a statistically rigorous approach to isolating aleatoric uncertainty, this work contributes to delivering climate information with appropriate expressions of confidence, supporting robust decision-making while acknowledging fundamental limits to predictability rather than conveying false precision.

\section*{Acknowledgments}
CG is supported by the DAAD program Konrad Zuse Schools of Excellence in Artificial Intelligence, sponsored by the Federal Ministry of Research, Technology and Space.

\section*{Conflict of interest}

The authors declare no potential conflict of interest.

\section*{Data Availability Statement}

The ERA5-Land reanalysis data that support the findings of this study are openly available in the Climate Data Store of Copernicus at https://doi.org/10.24381/cds.68d2bb30.
The CRCM5-LE data for the historical and RCP8.5 simulations are available from the ClimEx project https://www.climex-project.org/data-access/.


\bibliographystyle{unsrtnat}      
\bibliography{references}         

\appendix

\label{sec:appendix}
\section{Model Details and Estimation}
\label{sec:appendix_model_fitting}

\paragraph*{GAM Fitting}
We use the \texttt{bam()} function from the \texttt{mgcv} package, which is specifically designed for fitting GAMs on large datasets \citep{wood2011fast, wood2015GeneralizedAdditiveModels}. Estimation proceeds via penalized iteratively reweighted least squares (PIRLS), where spline coefficients are estimated jointly with smoothing penalties. Smoothing parameters $\lambda_k$ are selected by optimizing the (fast) restricted maximum likelihood (fREML) criterion, thus avoiding manual tuning.
The \texttt{bam()} implementation exploits sparse matrix methods and discrete approximations to achieve computational scalability for datasets of the size considered here, i.e., over 37 million observations.

\paragraph*{Choice of Tuning Parameters}
The basis dimensions ($k$) are chosen as follows:
\begin{itemize}
    \item Cyclic cubic splines are used for month ($k=12$) to enforce yearly periodicity.
    \item Penalized splines are used for elevation ($k=10$) and spatial effects, the latter specified as tensor product smooths ($k=25$) to allow anisotropic smoothing across longitude and latitude; smoothness is controlled via penalization to prevent overfitting.
    \item Random effects for member pairs are included via \texttt{bs="re"}.
\end{itemize}

\paragraph*{Interpretation}
Although the model does not explicitly specify a spatio-temporal covariance function, dependence is implicitly introduced through the smooth terms. 
The smooth terms ($f_k$) allow nearby observations (in space, time, or elevation) to inform each other, while still permitting flexible, non-linear patterns. 
From a probabilistic perspective, the model assumes conditional independence of observations given the covariates and smooth effects, such that residuals are treated as independent across $(s,t)$ and structured variation is captured through the mean. 
This provides a computationally efficient alternative to explicit covariance modeling while still capturing structured spatio-temporal variability.

\paragraph*{Validation}
The bias model in step 1 in \Cref{sec:val} is specified as:
\begin{equation}
\label{eq:appendix_gam_bias_reanalysis}
    \begin{split}
    \tilde{\delta}_{i,j}(s,t) =&~ \alpha_0 + \alpha_1 \cdot \text{year}_i 
    + g_1(\text{month}_i) + g_2(\text{elevation}_i) \\
    &+ \mathbbm{1}_{\text{season}_i} \cdot g_{\text{season}}(\text{lon}_i, \text{lat}_i) + u_j 
    + \tilde{\varepsilon}_{i,j}(s,t),
    \end{split}
\end{equation}
where $u_j \sim \mathcal{N}(0, \sigma_u^2)$ represents the random member-specific effect and   $\tilde{\varepsilon}_{i,j}(s,t)$ denotes the residual term.

\begin{table}[htb!]
\centering
\footnotesize
\renewcommand{\arraystretch}{1.2}
\setlength{\tabcolsep}{8pt}
\begin{tabular}{rllc}
i,j & \multicolumn{2}{c}{Member Pairs} & Family \\
\midrule
1,2 & kba & kbb & \multirow{5}{*}{Family 1} \\
3,4 & kbc & kbd & \\
5,6 & kbe & kbf & \\
7,8 & kbg & kbh & \\
9,10 & kbi & kbj & \\
\hline
11,12 & kbk & kbl & \multirow{5}{*}{Family 2} \\
13,14 & kbm & kbn & \\
15,16 & kbo & kbp & \\
17,18 & kbq & kbr & \\
19,20 & kbs & kbt & \\
\hline
21,22 & kbu & kbv & \multirow{5}{*}{Family 3} \\
23,24 & kbw & kbx & \\
25,26 & kby & kbz & \\
27,28 & kca & kcb & \\
29,30 & kcc & kcd & \\
\hline
31,32 & kce & kcf & \multirow{5}{*}{Family 4} \\
33,34 & kcg & kch & \\
35,36 & kci & kcj & \\
37,38 & kck & kcl & \\
39,40 & kcm & kcn & \\
\hline
41,42 & kco & kcp & \multirow{5}{*}{Family 5} \\
43,44 & kcq & kcr & \\
45,46 & kcs & kct & \\
47,48 & kcu & kcv & \\
49,50 & kcw & kcx & \\
\bottomrule
\end{tabular}
\caption{Member pairs used for calculating the squared differences.}
\label{tab:member_pairs}
\end{table}

\begin{table}[htb!]
\centering
\footnotesize
\begin{tabular}{clc}
j & Members & Family \\
\midrule
1 & kba & \multirow{5}{*}{Family 1} \\
4 & kbd & \\
5 & kbe & \\
6 & kbf & \\
10 & kbj & \\
\midrule
12 & kbl & \multirow{5}{*}{Family 2} \\
14 & kbn & \\
15 & kbo & \\
16 & kbp & \\
18 & kbr & \\
\midrule
22 & kbv & \multirow{5}{*}{Family 3} \\
26 & kbz & \\
27 & kca & \\
29 & kcc & \\
30 & kcd & \\
\midrule
34 & kch & \multirow{5}{*}{Family 4} \\
36 & kcj & \\
37 & kck & \\
38 & kcl & \\
40 & kcn & \\
\midrule
43 & kcq & \multirow{5}{*}{Family 5} \\
44 & kcr & \\
45 & kcs & \\
48 & kcv & \\
49 & kcw & \\
\bottomrule
\end{tabular}
\caption{Sampled members for calculating the validation.}
\label{tab:validation}
\end{table}

\begin{table}[t]
\centering
\begin{minipage}{0.95\textwidth}
\footnotesize
\begin{verbatim}
Family: Gamma 
Link function: log 

Formula:
squared_difference ~ te(rlon, rlat, bs = "ps", by = season) + 
    s(month, bs = "cc", k = 12) + s(topo, bs = "ps") + 
    s(member_difference, bs = "re") + year

Parametric coefficients:
             Estimate Std. Error 
(Intercept) 1.318e+00  8.885e-02  
year        1.490e-04  4.403e-05  


Approximate significance of smooth terms:
                             edf Ref.df       
te(rlon,rlat):seasonwinter 24.00 24.000 
te(rlon,rlat):seasonspring 23.97 23.999 
te(rlon,rlat):seasonsummer 23.96 23.999 
te(rlon,rlat):seasonautumn 23.99 24.000 
s(month)                   10.00 10.000 
s(topo)                     8.97  8.999 
s(member_difference)       23.96 24.000 

R-sq.(adj) =  0.168   Deviance explained = 27.3%
fREML = 8.4232e+07  Scale est. = 5.4075    n = 37224000

\end{verbatim}
\end{minipage}
\caption{GAM model output for \autoref{eq:gam_members}.}
\end{table}

\begin{table}[t]
\centering
\begin{minipage}{0.95\textwidth}
\footnotesize
\begin{verbatim}
Family: Gamma 
Link function: log 

Formula:
squared_res ~ te(rlon, rlat, bs = "ps", by = season) +  
    s(month, bs = "cc", k = 12) + s(topo, bs = "ps") + 
    s(member_difference, bs = "re") + year

Parametric coefficients:
              Estimate Std. Error 
(Intercept)  7.2357204  0.0834823  
year        -0.0028611  0.0000414 


Approximate significance of smooth terms:
                              edf Ref.df       
te(rlon,rlat):seasonwinter 23.994     24 
te(rlon,rlat):seasonspring 23.969     24 
te(rlon,rlat):seasonsummer 23.938     24 
te(rlon,rlat):seasonautumn 23.977     24 
s(month)                   10.000     10 
s(topo)                     8.986      9 
s(member_difference)       23.959     24 


R-sq.(adj) =  0.139   Deviance explained = 26.2%
fREML = 8.1937e+07  Scale est. = 4.7801    n = 37224000

\end{verbatim}
\end{minipage}
\caption{GAM model output for \autoref{eq:gam_variability_reanalysis}.}
\end{table}

\begin{table}[!htb]
\centering
\begin{minipage}{0.95\textwidth}
\begin{footnotesize}
\begin{verbatim}
Family: Gamma 
Link function: log 

Formula:
squared_difference ~ te(rlon, rlat, bs = "ps", by = season) + 
    s(month, bs = "cc", k = 12) + s(topo, bs = "ps") +  
    s(member_difference, bs = "re") + year

Parametric coefficients:
              Estimate Std. Error 
(Intercept)  5.623e+00  8.993e-02  
year        -2.008e-03  4.354e-05  


Approximate significance of smooth terms:
                              edf Ref.df 
te(rlon,rlat):seasonwinter 23.995 24.000 
te(rlon,rlat):seasonspring 23.968 23.999 
te(rlon,rlat):seasonsummer 23.874 23.988 
te(rlon,rlat):seasonautumn 23.984 24.000 
s(month)                   10.000 10.000 
s(topo)                     8.961  8.999 
s(member_difference)       23.969 24.000 


R-sq.(adj) =   0.18   Deviance explained =   27%
fREML = 8.3808e+07  Scale est. = 5.2856    n = 37224000

\end{verbatim}
\end{footnotesize}
\end{minipage}
\caption{GAM model output for future model.}
\end{table}

\end{document}